\documentclass{article}
\usepackage[T1]{fontenc}
\usepackage{spconf,amsmath,amssymb,graphicx,booktabs,tabularx,url}
\usepackage[table]{xcolor}
\newcolumntype{Y}{>{\centering\arraybackslash}X}
\title{AUDITING GENERATIVE AUDIO CALLS FOR KNOWN-TASK AUDIO-LLM EVALUATION}
\name{Mengzhe Geng\thanks{AI-use disclosure: OpenAI Codex (GPT-5) assisted with language editing and layout checking in Sections 1--6; the author verified all scientific content, data, results, and code.}}
\address{National Research Council Canada\\{\tt Mengzhe.Geng@nrc-cnrc.gc.ca}}

\begin{document}
\ninept
\maketitle

\begin{abstract}
Speech and audio LLMs are evaluated by comparing waveform predictions with predictions from an automatic speech recognition (ASR) transcript. For fixed closed-set tasks, this conflates acoustic evidence with the need to invoke a generative audio model. We estimate incremental call value with matched selectors sharing pre-call evidence. Each policy may retain the transcript label, use a local encoder, or invoke a generative model; matched control removes generative actions but preserves pre-call evidence and development selection. On VocalSound, transcript-only accuracy is 0.296, while supervised CLAP and WavLM controls reach 0.850 and 0.854 without calls. Full selector reaches 0.925 at 12.5\% calls versus 0.921 for matched No-call selector (difference 0.004; 95\% CI $[-0.025,0.033]$). Thus, results do not show a call gain after transcript and encoder evidence are available. Relevant quantity is incremental accuracy from allowing calls, not the waveform-transcript gap.
\end{abstract}


\section{Introduction}

Recent speech and audio LLMs take waveforms as input and return text decisions, descriptions, or dialogue responses \cite{speechgpt2023,audiopalm2023,salmonn2023,qwen2audio2024,miniomni2024,llamaomni2024,glm4voice2024,qwen25omni2025,mossaudio2026,stepaudio2025,audioflamingonext2026}. They are useful when automatic speech recognition (ASR) text discards information carried by the signal, such as vocalizations, emotion, speaker state, or overlapping speech \cite{speechlmsurvey2024,speechllmunderstanding2024}. A common evaluation compares an audio-LLM prediction with a transcript-only prediction. That test shows whether the waveform contains useful evidence, while leaving the system-design question open: did the system need a generative audio model to use that evidence?

In a deployed closed-set task, a system chooses among retaining a transcript-side label, scoring the waveform with a local encoder, and sending the waveform to a generative audio model. The last option sometimes improves decisions, but it also adds cost and, in hosted settings, increases exposure of speech data. We ask whether a generative audio call still changes the answer enough to justify its use after transcript and encoder evidence are available to a policy selected on the development set.

We answer this question with fixed development and holdout splits from the same dataset and matched controls. The evaluation covers Qwen2-Audio, Qwen2.5-Omni, MOSS-Audio, Contrastive Language--Audio Pretraining (CLAP), Audio Spectrogram Transformer (AST), and WavLM. The primary comparison matches Full and No-call on transcript, CLAP, and WavLM evidence. Results for the other endpoints and encoders are reported as secondary context. We evaluate transcript-first and encoder-first routes, compare matched Full and No-call selectors, report component costs, and compute paired confidence intervals. The scope is limited to closed-set label decisions. We do not evaluate open dialogue quality, instruction-following ability, or general audio reasoning.

\section{Related Work}

SUPERB evaluates reusable speech representations, and Dynamic-SUPERB extends this evaluation style to collaborative instruction-following speech tasks \cite{superb2021,dynamicsuperb2023}. Recent audio-language benchmarks add generative comprehension, voice-assistant behavior, hallucination, egocentric multi-talker speech, and speaker understanding \cite{airbench2024,audiobench2024,voicebench2024,halluaudio2026,smartglasses2026,asrllmstudy2024}. These benchmarks primarily measure model ability. Our evaluation instead fixes a known task and asks whether a generative audio call still changes the decision after transcript and encoder controls are included.

The closest prior diagnostic is Wu et al.'s study of whether speech-LLM speaker understanding can be explained from transcripts \cite{asrllmstudy2024}. Their Gaokao and identity-critical dialogue tests compare speech-LLM behavior with a transcript diagnostic; our test uses six predefined VocalSound classes and a per-example selector over transcript, local-encoder, and generative endpoints. Using fixed data splits, we pair Full with No-call under identical selector and development protocols and report call rate and component cost. This comparison tests the marginal value of a generative call, not general audio-LLM ability; Dynamic-SUPERB, AIR-Bench, and AudioBench remain broader capability benchmarks.

The framing also draws on calibration, misclassification detection, selective prediction, ASR confidence, and cost-aware cascades \cite{guo2017calibration,hendrycks2017baseline,geifman2017selective,oneata2021asrconfidence,frugalgpt2024}. In that literature, a defer-or-cascade policy decides when extra computation is warranted. Here, local audio encoders provide the first-stage evidence. We include CLAP, AST, and WavLM controls \cite{clap2023,ast2021,wavlm2022} and compare routed generative calls with selectors that see the same pre-call evidence but never invoke a generative audio model.

\section{Generative-Call Decision Test}

For utterance $x_i$, ASR produces transcript $y_i$, and a text LLM maps that transcript to label $\hat{c}^{\mathrm{text}}_i$. A generative audio model predicts $\hat{c}^{\mathrm{gen}}_i$ directly from the waveform. Given a target call budget $b$, a routing policy selects examples $\mathcal{R}_b$ for the generative model and returns
\begin{equation}
\hat{c}_i = \begin{cases}
\hat{c}^{\mathrm{gen}}_i, & i \in \mathcal{R}_b,\\
\hat{c}^{\mathrm{text}}_i, & i \notin \mathcal{R}_b,
\end{cases}
\end{equation}
where $\mathcal{R}_b=\{i:s_i\geq \tau_b\}$ and $\tau_b$ is chosen on development examples. The transcript routing score is
\begin{equation}
s_i = \rho_i^{\mathrm{text}}+\alpha u_i+\beta q_i+\gamma v_i,
\end{equation}
where $\mathcal{C}$ is the task label set, $\rho_i^{\mathrm{text}}$ is the development-estimated risk of the transcript label, $u_i$ is normalized Whisper uncertainty, $q_i$ is normalized text-LLM label-likelihood uncertainty, and $v_i$ is one when $\hat{c}^{\mathrm{text}}_i\notin\mathcal{C}$. We choose the weights and threshold on development examples only. We also test an encoder-first route: CLAP scores every waveform and sends low-margin examples to a generative audio model. Supervised CLAP, WavLM, and AST probes use labeled development examples; we treat them as non-generative controls for the fixed six-class setting.

		The primary endpoint comparison uses a full selector and a matched no-call selector. For each example--action pair, we fit an L2-regularized logistic model that predicts whether the action is correct. Features include action confidence, transcript uncertainty and length, encoder and call indicators, action cost, and action identity. CLAP and WavLM development predictions are out-of-fold. We choose $\lambda$ on development accuracy, breaking ties toward fewer generative calls. We create the No-call control by removing the Qwen2-Audio and Qwen2.5-Omni actions while leaving training and selection unchanged. The Full--No-call difference therefore estimates the incremental value of permitting a generative action with the same pre-call evidence.

		Qwen2-Audio and Qwen2.5-Omni do not provide comparable token-level confidence, so we use a constant confidence feature for those endpoints. The comparison does not test calibrated endpoint confidence and is conditional on the endpoint predictions. Post-call agreement and stacked-fusion analyses are excluded from the primary comparison because they use predictions available only after invocation.

		Before evaluating the holdout examples, we fix the parser, label normalization, features, weights, thresholds, and $\lambda$ using development examples, then apply them unchanged to the holdout. Paired intervals use 10,000 paired bootstrap resamples over examples, McNemar $p$ values are exact, point intervals use Wilson intervals, and the eight prespecified paired tests use Holm correction. In a 10-seed diagnostic, we vary the CLAP/WavLM fold assignments and refit the selector while holding generative-model predictions fixed.

	We separate cost from accuracy. For a transcript-first route, the measured sequential component cost is
\begin{equation}
C_i^{\mathrm{route}} = C_i^{\mathrm{asr}} + C_i^{\mathrm{text}} + h_i C_i^{\mathrm{conf}} + r_i C_i^{\mathrm{gen}},
\end{equation}
	where $h_i$ marks optional transcript-confidence scoring and $r_i$ marks a generative audio call. For the primary selectors, transcript features and both local-encoder feature sets are available to every candidate action. Their total cost before routing is therefore
\begin{equation}
\begin{aligned}
C_i^{\mathrm{pre}} &= C_i^{\mathrm{asr}} + C_i^{\mathrm{text}}\\
                    &\quad + C_i^{\mathrm{clap}} + C_i^{\mathrm{wavlm}},\\
C_i^{\mathrm{Full}} &= C_i^{\mathrm{pre}} + r_i C_i^{\mathrm{gen}}.
\end{aligned}
\end{equation}
where $C_i^{\mathrm{No\text{-}call}}=C_i^{\mathrm{pre}}$. For a CLAP-first policy, the analogous cost is $C_i^{\mathrm{clap}} + r_i C_i^{\mathrm{gen}}$: every waveform reaches CLAP, but only routed examples reach the generative audio model. We report aggregate component runtime over the holdout, summing measured times for the components required by each route. These totals exclude model startup, batching, concurrency, network overhead, probe fitting, and optimized serving effects. This accounting reports component runtimes and the reduction in generative calls only. Privacy exposure and end-to-end serving latency require separate measurements because hosted interfaces may receive transcripts, embeddings, prompts, or routed waveforms.

\section{Evaluation}

		The main benchmark is Dynamic-SUPERB VocalSound, a six-way human-vocalization task with chance accuracy 0.167 and labels for laughter, sigh, cough, sneeze, sniff, and throat clearing \cite{vocalsound2022}. Dataset indices 0--479 form the development split, and indices 480--719 form the same-dataset holdout. The partitions contain no repeated audio files, but the dataset provides no speaker or session identifiers. Separation by speaker, session, recording period, or licensing source is therefore unknown.

		The transcript pipeline uses \texttt{openai/}\allowbreak\texttt{whisper-large-v3-}\allowbreak\texttt{turbo} for automatic speech recognition. It uses \texttt{Qwen/}\allowbreak\texttt{Qwen2.5-}\allowbreak\texttt{1.5B-}\allowbreak\texttt{Instruct} for transcript-side label decisions. We also evaluate three external checks. ESC-50 Animals ($n=200$) is a ten-way environmental-sound task with labels dog, rooster, pig, cow, frog, cat, hen, insects, sheep, and crow \cite{esc502015}. The five-emotion AudioBench paralinguistic question-answering (PQA) subset is drawn from the English portion of Emotional Speech Dataset (ESD), with labels angry, happy, neutral, sad, and surprised, development $n=90$, and holdout $n=120$ \cite{audiobench2024,esd2022}. Dynamic-SUPERB SpeechTextMatching LibriSpeech-TestClean (STM; $n=100$) is evaluated over LibriSpeech \cite{dynamicsuperb2023,librispeech2015}. 

		VocalSound first confirms that transcripts are insufficient: the transcript-only system reaches 0.296 accuracy. Audio models are stronger: Qwen2.5-Omni reaches 0.883, Qwen2-Audio 0.838, and MOSS-Audio 0.808. Zero-shot CLAP reaches 0.762 and AudioSet AST 0.404.
		Encoder controls trained on labeled development data narrow the gap further. CLAP class-bias tuning reaches 0.792. A CLAP-score probe reaches 0.796 accuracy and 0.784 macro F1; CLAP-embedding and AST-embedding probes reach 0.850/0.849 and 0.817/0.813, respectively. WavLM-base+ reaches 0.854/0.853. A CLAP--WavLM correctness oracle reaches 0.938, and a joint CLAP--WavLM ridge probe reaches 0.896/0.892.

\begin{figure*}[t]
\centering
\includegraphics[width=\textwidth]{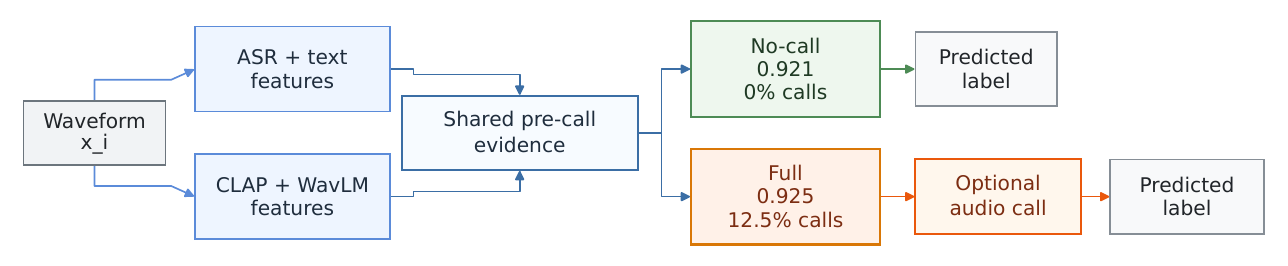}
\caption{Decisions before and after a generative-model call in the primary VocalSound comparison. ASR/text and CLAP/WavLM features form shared pre-call evidence. No-call stops before a generative endpoint; Full may issue an optional call. The annotations are holdout summaries (0.921 at 0\% calls and 0.925 at 12.5\% calls), not a deployment graph or latency estimate.}
\label{fig:callboundary}
\end{figure*}

\begin{figure*}[t]
\centering
\includegraphics[width=\textwidth]{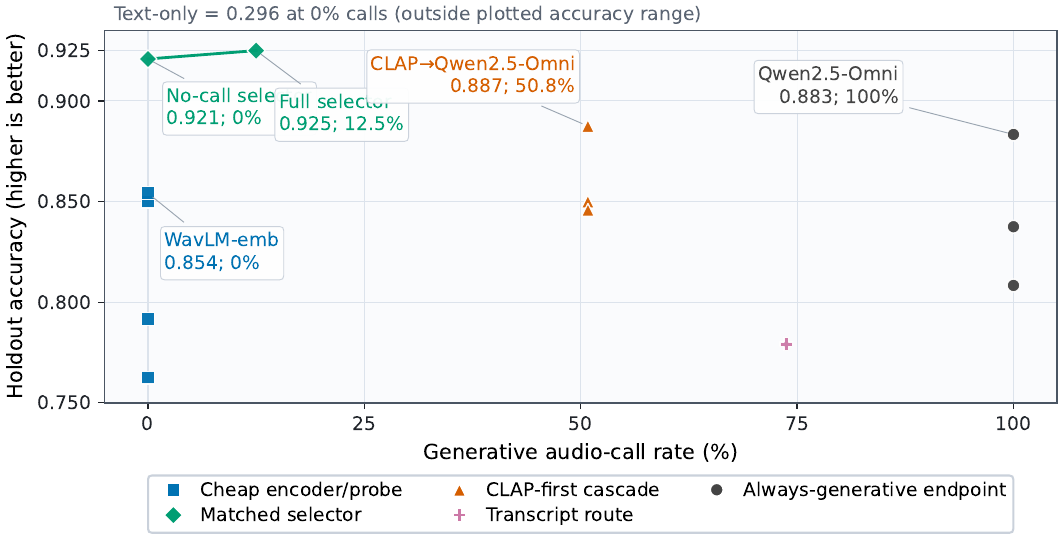}
\caption{VocalSound holdout accuracy versus generative-call rate for representative policies. The y-axis is explicitly cropped to the competitive range (0.75--0.935); the transcript-only baseline (0.296 at 0\% calls) is annotated above the panel. The x-axis counts only generative audio-model calls; no-endpoint controls still process transcripts or local audio features. The full selector is highest in this summary, but its paired difference from the matched no-call selector has a 95\% CI that includes zero.}
\label{fig:callrate}
\end{figure*}

		Figure~\ref{fig:callboundary} shows why the remaining gap is small: both policies already use transcript, CLAP, and WavLM evidence. The optional endpoint can therefore affect only the remaining cases after this pre-call evidence is available. This changes the comparison from transcript-versus-audio accuracy to the marginal value of a generative audio call. The representative call-rate/accuracy summary is shown in \mbox{Fig.~\ref{fig:callrate}}; representative policy costs and paired accuracy contrasts are listed in \mbox{Table~\ref{tab:policycost}}.

\begin{table}[!b]
\caption{VocalSound holdout summary ($n=240$). Generative-call rate is the fraction of the 240 holdout examples sent to a generative audio model. Component times are aggregate runtime, not per-example latency or end-to-end serving cost. For the selector rows, the total includes mandatory ASR, transcript-side, CLAP, and WavLM components; the endpoint column is conditional on selected calls. The primary paired contrast is Full--No-call; each remaining contrast is labeled with its reference policy and is secondary or exploratory.}
\label{tab:policycost}
\vspace{3pt}
\centering
\setlength{\tabcolsep}{2.4pt}
\renewcommand{\arraystretch}{1.06}
\fontsize{9}{10.2}\selectfont
\begin{tabularx}{\columnwidth}{@{}lYYYY@{}}
\toprule
\multicolumn{5}{@{}l}{\textbf{Policy results and runtime}}\\
\midrule
\textbf{Policy} & \textbf{Accuracy ($\uparrow$)} & \textbf{Call rate (\%)} & \textbf{Total runtime (s)} & \textbf{Gen. runtime (s)} \\
\midrule
Text & 0.296 & 0.0\% & 34.4 & 0.0 \\
WavLM-emb & 0.854 & 0.0\% & 2.5 & 0.0 \\
Qwen2-Audio & 0.838 & 100.0\% & 31.5 & 31.5 \\
\shortstack[l]{CLAP$\to$Qwen2.5-\\Omni} & 0.887 & 50.8\% & 30.5 & 21.7 \\
\shortstack[l]{No-call\\selector} & 0.921 & 0.0\% & 45.8 & 0.0 \\
\rowcolor{gray!12} \shortstack[l]{Full\\selector} & \textbf{0.925} & 12.5\% & 51.4 & 5.6 \\
\bottomrule
\end{tabularx}
\vspace{2pt}

\begin{tabularx}{\columnwidth}{@{}lYY@{}}
\toprule
\multicolumn{3}{@{}l}{\textbf{Paired accuracy differences}}\\
\midrule
\textbf{Contrast} & \textbf{Accuracy difference} & \textbf{95\% CI} \\
\midrule
\shortstack[l]{CLAP$\to$Qwen2.5-\\Omni--WavLM (secondary)} & 0.033 & $[-0.017,0.083]$ \\
Full--WavLM (exploratory) & 0.071 & $[0.029,0.117]$ \\
Full--No-call (primary) & 0.004 & $[-0.025,0.033]$ \\
\bottomrule
\end{tabularx}
\end{table}

		The primary matched comparison is Full--No-call: the full selector reaches 0.925 accuracy at 12.5\% generative calls, while the matched no-call selector reaches 0.921 ($\Delta=0.004$, 95\% CI $[-0.025,0.033]$). The matched policies differ in correctness on only 13 of the 240 holdout examples: Full is correct where No-call is wrong in 7 cases, and wrong where No-call is correct in 6. Because the interval includes zero, the VocalSound results do not show a generative-call gain after matched pre-call evidence and development selection.

		For context, an unattainable oracle that chooses the correct available action reaches 0.975 with 2.9\% endpoint calls; this is an upper bound on the available action set, not a deployable policy. The Full--WavLM result is exploratory because it compares different selector families. Its larger difference (0.071, 95\% CI $[0.029,0.117]$) cannot replace the matched primary contrast. CLAP-first Qwen2-Audio reaches 0.850 accuracy with 50.8\% generative calls, matching the supervised CLAP-embedding probe but with lower macro F1, the unweighted class average (0.841 versus 0.849), and higher call cost.

		The remaining VocalSound diagnostics are secondary. CLAP-first Qwen2-Audio and CLAP-first MOSS-Audio are slightly below WavLM-emb. The best routed policy sends low-margin CLAP examples to Qwen2.5-Omni and reaches 0.887 with 50.8\% generative calls. Its paired advantage over WavLM-emb is not detected ($\Delta=0.033$, 95\% CI $[-0.017,0.083]$, McNemar $p=0.268$). Across prompt variants, 50\% CLAP-first Qwen2-Audio accuracy ranges from 0.792 to 0.850. This result depends on label wording and encoder access.

		Cost accounting reports aggregate component runtime over the $n=240$ holdout examples. It is not per-example latency or end-to-end serving latency. Because the selector uses transcript and both encoder feature sets before choosing an action, the No-call total including all required pre-call components is 45.8~s and the full-selector total is 51.4~s. The selected generative component contributes 0.0~s and 5.6~s, respectively. Totals that omit mandatory pre-call selector components are not reported. The WavLM total is allocated uniformly because per-record WavLM latency was not measured.

		Across the prespecified paired tests, the exploratory Full--WavLM contrast is the only raw comparison surviving Holm correction ($p_{\mathrm{adj}}=0.027$); this cross-family result does not establish a generative-call gain over the matched no-call control. Post-call agreement and stacked-fusion checks are excluded from the primary evidence about the value of a generative call because they require endpoint predictions after invocation.

		The external checks point in the same direction. ESC-50 Animals confirms the value of acoustic evidence over text alone (CLAP 0.895, Qwen2-Audio 0.785, transcript-only 0.140). On an earlier ESD split, Qwen2-Audio reaches 0.425 accuracy and 0.341 macro F1, while WavLM reaches 0.475 accuracy and 0.444 macro F1. The raw comparison between Full and No-call gives 0.433 versus 0.425 ($\Delta=0.008$, CI $[-0.067,0.083]$). An agreement-feature selector reaches 0.442 accuracy, compared with 0.475 for the matched No-call selector. A stacked-fusion analysis reaches 0.550 but uses endpoint predictions and is post-call.

		A separate speaker-disjoint ESD evaluation uses five development and five holdout speakers, with 125 balanced examples per partition and no shared speakers or source recordings. Its holdout accuracies are 0.200 for Qwen2-Audio, 0.272 for Qwen2.5-Omni, 0.200 for CLAP, and 0.184 for the transcript control. Both partitions come from ESD, so this evaluation does not test cross-dataset generalization. STM reverses the ordering, with transcript accuracy 0.960 versus Qwen2-Audio 0.850, so call decisions are task-specific.

\section{Discussion and Limitations}

	The evaluation inherits the biases of ASR systems, audio encoders, and LLMs on accents, dialects, disordered speech, code-switching, and low-resource languages.

		Reducing generative calls does not eliminate exposure of audio-derived data. CLAP-first and Full still process transcripts and encoder features, while hosted interfaces may also receive routed waveforms or intermediate representations. The partitions contain no repeated audio files, but the dataset lacks speaker, session, and broader source metadata. Repeated reuse may encourage test-set overfitting and weaken its value as an independent holdout.

		A separate speaker-disjoint ESD evaluation addresses speaker and source-recording separation, but remains a small within-dataset robustness check. Its near-baseline holdout results do not establish that the generative endpoint adds value on other datasets or tasks. The conclusion is also task-specific. Supervised CLAP/WavLM embeddings match or exceed routed generative calls here. These findings may not extend to richer reasoning tasks, unknown task mixtures, few-shot adaptation, settings with calibrated generative log-probabilities, or throughput-optimized serving. Each setting requires separate evaluation. Deployment use requires consent, privacy review, fairness analysis, and strict data minimization.

\section{Conclusion}

An audio-vs-transcript gain is insufficient evidence that a generative audio call was necessary. On VocalSound, transcript-only accuracy is 0.296 for non-speech vocalization classification. Supervised encoders are strong when development labels exist, and the best no-call selector nearly ties a full selector with access to generative audio models. Future evaluations should separate decisions made before a generative-model call from changes made after the call. They should report what is decided from transcripts and local encoders, and what changes only after the waveform reaches the generative model. They should also test whether a pre-call signal can identify cases in which a call corrects, rather than introduces, an error.

\bibliographystyle{IEEEbib}
\bibliography{refs}

\end{document}